\documentclass[english]{article}
\usepackage[T1]{fontenc}
\usepackage{graphicx}
\usepackage[latin1]{inputenc}
\usepackage{geometry}
\usepackage{amsmath}
\usepackage{color}
\usepackage{amssymb}

\makeatletter
\AtBeginDocument{
  
}

\usepackage{babel}
\makeatother
\begin{document}
\begin{center}\textbf{\large Higher order antibunching in intermediate
states}{\LARGE }\\
{\LARGE ~}\\
\par\end{center}{\LARGE \par}

\begin{center}\textcolor{black}{\large Amit Verma}%
\footnote{amit.verma@jiit.ac.in%
}\textcolor{black}{\large , Navneet K Sharma}%
\footnote{navneetk.sharma@jiit.ac.in%
} \textcolor{black}{\large and Anirban Pathak}\textcolor{black}{}%
\footnote{\textcolor{black}{anirban.pathak@jiit.ac.in}%
}\par\end{center}

\begin{center}\textcolor{black}{Department of Physics, JIIT University,
A-10, Sectror-62, Noida, UP-201307, India.}\\
\par\end{center}

\begin{abstract}
\textcolor{black}{\normalsize Since the introduction of binomial
state as an intermediate state, different intermediate states have
been proposed. Different nonclassical effects have also been
reported in these intermediate states. But till now higher order
antibunching or higher order subpoissonian photon statistics is
predicted only in one type of intermediate state, namely shadowed
negative binomial state. Recently we have shown the existence of
higher order antibunching in some simple nonlinear optical processes
to establish that higher order antibunching is not a rare phenomenon
(}J. Phys. B 39 (2006) 1137\textcolor{black}{\normalsize ). To
establish our earlier claim further, here we have shown that the
higher order antibunching can be seen in different intermediate
states, such as binomial state, reciprocal binomial state,
hypergeometric state, generalized binomial state, negative binomial
state and photon added coherent state. We have studied the
possibility of observing the higher order subpoissonian photon
statistics in different limits of intermediate states. The effect of
different control parameters have also been studied in this
connection and it has been shown that the depth of nonclassicality
can be tuned by controlling various physical
parameters.}{\normalsize
\par}
\end{abstract}

\section{\textcolor{black}{\normalsize Introduction}}

An intermediate state is a quantum state which reduces to two or
more distinguishably different states (normally, distinguishable in
terms of photon number distribution) in different limits. In 1985,
such a state was first time introduced by Stoler \emph{et al.}
{[}\ref{the:D-Stoler,-B}]. To be precise, they introduced Binomial
state (BS) as a state which is intermediate between the most
nonclassical number state $|n\rangle$ and the most classical
coherent state $|\alpha\rangle$. They defined BS as \begin{equation}
\begin{array}{lr}
|p,M\rangle=\sum_{n=0}^{M} & B_{n}^{M}\end{array}|n\rangle=\sum_{n=0}^{M}\sqrt{^{M}C_{n}p^{n}(1-p)^{M-n}}|n\rangle\,\,\,\:0\leq p\leq1.\label{eq:binomial1}\end{equation}
This state%
\footnote{The state is named as binomial state because the photon
number distribution associated with this state
$\left(i.e.\,|B_{n}^{M}|^{2}\right)$ is
simply a binomial distribution.%
} is called intermediate state as it reduces to number state in the
limit $p\rightarrow0$ and $p\rightarrow1$ (as
$|0,M\rangle=|0\rangle$ and $|1,M\rangle=|M\rangle$) and in the
limit of $M\rightarrow\infty,\, p\rightarrow1$, where $\alpha$ is a
real constant, it reduces to a coherent state with real amplitude.
Since the introduction of BS as an intermediate state, it was always
been of interest to quantum optics, nonlinear optics, atomic physics
and molecular physics community. Consequently, different properties
of binomial states have been studied
{[}\ref{the:VIdiella-Barraco}-\ref{the:OEBS}]. In these studies it
has been observed that the nonclassical phenomena (such as,
antibunching, squeezing and higher order squeezing) can be seen in
BS. This trend of search for nonclassicality in Binomial state,
continued in nineties. In one hand, several versions of generalized
BS have been proposed
{[}\ref{the:hong-chenfu-genralized-BS}-\ref{the:Hong-yi-Fan-generalized-BS}]
and in the other hand, people went beyond binomial states and
proposed several other form of intermediate states (such as,
\textcolor{black}{excited binomial state {[}\ref{the:EBS}], odd
excited binomial state {[}\ref{the:OEBS}], hypergeometric state
{[}\ref{the:hypergeomtric}], negative hypergeometric state
{[}\ref{the:negativehyper geometric}], reciprocal binomial state
{[}\ref{the:reciprocalBS}], shadowed state
{[}\ref{the:CTLee(shaddowed)}], shadowed negative binomial state
{[}\ref{the:ranjani and ctlee}] and photon added coherent state
{[}\ref{the:agarwal-photonadded}] etc.).} The studies in the
nineties were mainly limited to theoretical predictions but the
recent developments in the experimental techniques made it possible
to verify some of those theoretical predictions. For example, we can
note that, as early as in 1991 Agarwal and Tara
{[}\ref{the:agarwal-photonadded}] had introduced photon added
coherent state as \begin{equation} |\alpha,m\rangle=\frac{a^{\dagger
m}|\alpha\rangle}{\langle a^{m}a^{\dagger
m}\rangle},\label{eq:photonadded1}\end{equation} (where $m$ is an
integer and $|\alpha\rangle$ is coherent state) but the experimental
generation of the state has happened only in recent past when
Zavatta, Viciani and Bellini {[}\ref{the:photonadded-experiment}]
succeed to produce it in 2004. It is easy to observe that
(\ref{eq:photonadded1}) represents an intermediate state, since it
reduces to coherent state in the limit $m\rightarrow0$ and to number
state in the limit $\alpha\rightarrow0$. \textcolor{black}{This
state can be viewed as a coherent state in which additional $m$
photon are added. The photon number distribution of all the above
mentioned states are different but all these states belong to a
common family of states called intermediate state.} It has also been
found that most of these intermediate states show antibunching,
squeezing, higher order squeezing, subpoissonian photon statistics
etc. but higher order antibunching has been reported only in
shadowed negative binomial state {[}\ref{the:ranjani and ctlee}].
Inspired by these observations, many schemes to generate
intermediate states have been proposed in recent past
{[}\ref{the:photonadded-experiment}-\ref{the:R LOFranco2}].

\textcolor{black}{The reason behind the study of nonclassical properties
of intermediate states lies in the fact that the most of the interesting
recent developments in quantum optics have arisen through the nonclassical
properties of the radiation field only. For example, antibunching
and squeezing, which do not have any classical analogue {[}\ref{elements of quantum optics}-\ref{hbt}],
have extensively been studied in last thirty years. But the majority
of these studies are focused on lowest order nonclassical effects.
Higher order extensions of these nonclassical states have only been
introduced in recent past {[}\ref{hong}-\ref{lee2}].  Among these
higher order nonclassical effects, higher order squeezing has already
been studied in detail {[}\ref{hong}, \ref{hong2}, \ref{hillery},
\ref{giri}] but the higher order antibunching (HOA) is not yet studied
rigourously. }

\textcolor{black}{The idea of HOA was introduced by Lee in a
pioneering paper {[}\ref{lee1}] in 1990, since then it has been
predicted in two photon coherent state {[}\ref{lee1}], shadowed
negative binomial state {[}\ref{the:ranjani and ctlee}], trio
coherent state {[}\ref{ba an}] and in the interaction of intense
laser beam with an inversion symmetric third order nonlinear medium
{[}\ref{the:martin}]. From the fact that in first 15 years after its
introduction, HOA was reported only in some particular cases, HOA
appeared to be a very rare phenomenon. But recently we have shown
that the HOA is not a rare phenomenon {[}\ref{the:pathak-jphysB}]
and it can be seen in simple optical processes} like six wave mixing
process, four wave mixing process and second harmonic
generation\textcolor{black}{. To establish that further, here we
have shown the existence of HOA in different intermediate states,
namely, binomial state, reciprocal binomial state, photon added
coherent state, hypergeometric state, Roy-Roy generalized binomial
state and negative binomial state. }

\textcolor{black}{The present work is motivated by the recent
experimental observation of intermediate state
{[}\ref{the:photonadded-experiment}], theoretical observation of
possibility of observing HOA in some simple optical systems
{[}\ref{the:pathak-jphysB}] and the fact that the intermediate
states, which frequently show different kind of nonclassicality,
form a big family of quantum state. But till now HOA has been
predicted only in one member  (Shadowed negative binomial state) of
such a big family of quantum states {[}\ref{the:ranjani and ctlee}].
Motivated by these facts the present work aims to study the
possibility of HOA in all the popularly known intermediate states.
The theoretical predictions of the present study can be
experimentally verified with the help of various intermediate state
generation schemes and homodyne experiment, since the criteria for
HOA appears in terms of factorial moment, which can be measured by
using homodyne photon counting experiments
{[}\ref{the:Bachor-H-A,}-\ref{the:singh2}]. In the next section we
have briefly described the criteria of HOA. In section 3 we have
shown that the HOA of any arbitrary order can be seen in BS. In
section 4, Roy-Roy generalized binomial state
{[}\ref{the:broy&proy-generalized-bs}] is studied and existence of
HOA is predicted. Calculational details and methodology have been
shown only in algebraically simple cases which are described in
section 3 and 4. Section 5 is divided in several subsections and we
have followed the same procedure and have studied the possibilities
of observing HOA in different intermediate states namely, reciprocal
binomial state, negative binomial state, hypergeometric states and
photon added coherent state. One subsection is dedicated for the
discussion of one particular intermediate state. Finally section 6
is dedicated to conclusions. }

\section{\textcolor{black}{\normalsize Criteria of HOA}}
\textcolor{black}{The criterion of HOA is expressed in terms of
higher order factorial moments of number operator. There exist
several criterion for the same which are essentially equivalent.
Initially, using the negativity of P function {[}\ref{elements of
quantum optics}], Lee introduced the criterion for HOA as }

\textcolor{black}{\begin{equation} R(l,m)=\frac{\left\langle
N_{x}^{(l+1)}\right\rangle \left\langle N_{x}^{(m-1)}\right\rangle
}{\left\langle N_{x}^{(l)}\right\rangle \left\langle
N_{x}^{(m)}\right\rangle }-1<0,\label{eq:ho3}\end{equation} where
$N$ is the usual number operator, $\left\langle N^{(i)}\right\rangle
=\left\langle N(N-1)...(N-i+1)\right\rangle $ is the $ith$ factorial
moment of number operator, $\left\langle \right\rangle $ denotes the
quantum average, $l$ and $m$ are integers satisfying the conditions
$1\leq m\leq l$ and the subscript $x$ denotes a particular mode. Ba
An {[}\ref{ba an}] choose $m=1$ and reduced the criterion of $l$th
order antibunching to \begin{equation} A_{x,l}=\frac{\left\langle
N_{x}^{(l+1)}\right\rangle }{\left\langle N_{x}^{(l)}\right\rangle
\left\langle N_{x}\right\rangle }-1<0\label{eq:bhuta1}\end{equation}
or, \begin{equation} \left\langle N_{x}^{(l+1)}\right\rangle
<\left\langle N_{x}^{(l)}\right\rangle \left\langle
N_{x}\right\rangle .\label{eq:ba an (cond)}\end{equation}
Physically, a state which is antibunched in $l$th order has to be
antibunched in $(l-1)th$ order. Therefore, we can further simplify
(\ref{eq:ba an (cond)}) as \begin{equation} \left\langle
N_{x}^{(l+1)}\right\rangle <\left\langle N_{x}^{(l)}\right\rangle
\left\langle N_{x}\right\rangle <\left\langle
N_{x}^{(l-1)}\right\rangle \left\langle N_{x}\right\rangle
^{2}<\left\langle N_{x}^{(l-2)}\right\rangle \left\langle
N_{x}\right\rangle ^{3}<...<\left\langle N_{x}\right\rangle
^{l+1}\label{eq:ineq}\end{equation} and obtain the condition for
$l-th$ order antibunching as \begin{equation} d(l)=\left\langle
N_{x}^{(l+1)}\right\rangle -\left\langle N_{x}\right\rangle
^{l+1}<0.\label{eq:ho21}\end{equation} This simplified criterion
(\ref{eq:ho21}) coincides exactly with the physical criterion of HOA
introduced by Pathak and Garica {[}\ref{the:martin}] and the
criterion of Erenso, Vyas and Singh {[}\ref{the:Erenso}], recently
Vogel has reported a class of nonclassicality conditions based on
higher order factorial moments {[}\ref{the:vogel}]. All these
criteria essentially lead to same kind of nonclassicality which
belong to the class of strong nonclassicality according to the
classification scheme of Arvind} \textcolor{black}{\emph{et al}}
\textcolor{black}{{[}\ref{the:Arvind}]. Here we can note that
$d(l)=0$ and $d(l)>0$ corresponds to higher order coherence and
higher order bunching (many photon bunching) respectively.}
Actually, $\left\langle a^{\dagger l}a^{l}\right\rangle
=\left\langle N^{(l)}\right\rangle $ is a measure of the probability
of observing $l$ photons of the same mode at a particular point in
space time coordinate. Therefore the physical meaning of
inequalities (\ref{eq:ineq}) is that the probability of detection of
single photon pulse is greater than that of a two photon in a bunch
and that is greater than the probability of detection of three
photon in a bunch and so on. This is exactly the characteristic that
is required in a probabilistic single photon source used in quantum
cryptography. In other words all the probabilistic single photon
sources used in quantum cryptography should satisfy the criteria
(\ref{eq:ho21}) of HOA {[}\ref{the:IJP}].

\section{Binomial State}

Binomial state is originally defined as (\ref{eq:binomial1}), from
which it is straight forward to see that \begin{equation}
\begin{array}{lcl}
a|p,M\rangle & = & \sum_{n=0}^{M}\left\{ \frac{M!}{(n-1)!(M-n)!}p^{n}(1-p)^{M-n}\right\} ^{\frac{1}{2}}|n-1>\\
 & = & \sum_{l=0}^{M-1}\left\{ \frac{M(M-1)!}{l!(M-1-l)!}p^{l+1}(1-p)^{M-1-l}\right\} ^{\frac{1}{2}}|l>\,\,\,\,\,\,\,\,(assuming\, n-1=l)\\
 & = & [Mp]^{\frac{1}{2}}\sum_{l=0}^{M-1}\left\{ \frac{(M-1)!}{l!(M-1-l)!}p^{l}(1-p)^{M-1-l}\right\} ^{\frac{1}{2}}|l>\\
 & = & [Mp]^{\frac{1}{2}}|p,M-1\rangle.\end{array}\label{eq:eigen1}\end{equation}
 Similarly, we can write, \begin{equation}
\begin{array}{lcl}
a^{2}|p,M\rangle & = & [M(M-1)p^{2}]^{\frac{1}{2}}|p,M-2\rangle\\
a^{3}|p,M\rangle & = & [M(M-1)(M-2)p^{3}]^{\frac{1}{2}}|p,M-3\rangle\\
\vdots & \vdots & \vdots\\
a^{l}|p,M\rangle & = & [M(M-1)....(M-l+1)p^{l}]^{\frac{1}{2}}|p,M-l\rangle\\
 & = & \left[\frac{M!}{(M-l)!}p^{l}\right]^{\frac{1}{2}}|M-l,p\rangle.\end{array}\label{eq:eigen2}\end{equation}
Therefore, \begin{equation} \langle M,p|a^{\dagger l}=\langle
M-l,p|\left[\frac{M!}{(M-l)!}p^{l}\right]^{\frac{1}{2}}\label{eq:eigen3}\end{equation}
and consequently, \begin{equation} \langle
M,p|n^{(l)}|p,M\rangle=\langle M,p|a^{\dagger
l}a^{l}|p,M\rangle=\left[\frac{M!}{(M-l)!}p^{l}\right].\label{eq:expectation1}\end{equation}
Now substituting (\ref{eq:expectation1}) in equation (\ref{eq:ho21})
we obtain the condition for $lth$ order antibunching as
\begin{equation}
d(l)=\left[\frac{M!}{(M-l-1)!}p^{l+1}\right]-[Mp]^{l+1}<0\label{eq:dl-binomial}\end{equation}
or, \begin{equation}
(M-1)(M-2)....(M-l)<M^{l}\label{eq:hoasatisfied}\end{equation} which
is always satisfied for any $M>l$ and both $M$ and $l$ are positive
(since every term in left is $<M$). As $M$ is the number of photons
present in the field and $d(l)$ is a measure of correlation among
$(l+1)$ photons, therefore $M\geq(l+1)$ or $M>l$. Consequently, a
binomial state always shows HOA and the highest possible order of
antibunching that can be seen in a binomial state is equal to $M-1$,
where $M$ is the number of photon present in the field. From
(\ref{eq:dl-binomial}) it is straight forward to see that the number
state is always higher order antibunched and in the other extreme
limit (when $p\rightarrow1,\, M\rightarrow\infty$ and the BS reduces
to coherent state) $d(l)=0,$ which is consistent with the physical
expectation.

\section{Generalized Binomial State}

We have already mentioned that there are different form of
generalized binomial states
{[}\ref{the:hong-chenfu-genralized-BS}-\ref{the:Hong-yi-Fan-generalized-BS}].
For the present study we have chosen generalized binomial state
introduced by Roy and Roy {[}\ref{the:broy&proy-generalized-bs}].
They have introduced the generalized binomial state (GBS) as
\begin{equation}
|N,\alpha,\beta\rangle=\sum_{n=o}^{N}\sqrt{\omega(n,N,\alpha,\beta)}|n\rangle\label{eq:gen-bino1}\end{equation}
where, \begin{equation}
\omega(n,N,\alpha,\beta)=\frac{N!}{(\alpha+\beta+2)_{N}}\frac{(\alpha+1)_{n}(\beta+1)_{N-n}}{n!(N-n)!}\label{eq:gen-bino2}\end{equation}
 with $\alpha,\beta>-1$, $n=0,1,....,N$, and \begin{equation}
\begin{array}{lcr}
(a)_{0}=1 & \,\,\,\,\,\,\, &
(a)_{n}=a(a+1)....(a+n-1)\end{array}.\label{eq:gen-bino3}\end{equation}
This intermediate state reduces to vacuum state, number state,
coherent state, binomial state and negative binomial state in
different limits of $\alpha$, $\beta$ and $N$. In order to obtain an
analytic expression of $d(l)$ for this particular generalized
binomial state we need to prove following useful identity:

\textbf{Identity1:} \begin{equation}
a(a+1)_{n}=(a)_{n+1}\label{eq:identity 1}\end{equation}

Proof: Using (\ref{eq:gen-bino3}) we can write \[
(a+1)_{n}=(a+1)...(a+n)=\frac{a(a+1)...(a+n)}{a}=\frac{(a)_{n+1}}{a}.\]
Therefore, \[ a(a+1)_{n}=(a)_{n+1}.\]

Now it is easy to see that the above identity (\ref{eq:identity 1})
yields the following useful relations: \begin{equation}
(\alpha+1)_{l+1}=(\alpha+1)(\alpha+2)_{l}\label{eq:relation1}\end{equation}
 and \begin{equation}
(\alpha+\beta+2)_{N}=(\alpha+\beta+2)(\alpha+\beta+3)_{N-1}=(\alpha+\beta+2)(\alpha+2+\beta+1)_{N-1}.\label{eq:relation2}\end{equation}
Using (\ref{eq:gen-bino1}) and (\ref{eq:gen-bino2}) we can obtain
\begin{equation}
\begin{array}{lcl}
a|N,\alpha,\beta\rangle & = & \sum_{n=0}^{N}\left\{ \frac{N!}{(\alpha+\beta+2)_{N}}\frac{(\alpha+1)_{n}}{(n-1)!}\frac{(\beta+1)_{N-n}}{(N-n)!}\right\} ^{\frac{1}{2}}|n-1\rangle\\
 & = & \sum_{l=0}^{N-1}\left\{ \frac{N(N-1)!}{(\alpha+\beta+2)_{N}}\frac{(\alpha+1)_{l+1}}{l!}\frac{(\beta+1)_{N-1-l}}{(N-1-l)!}\right\} ^{\frac{1}{2}}|l\rangle\end{array},\label{eq:relation 3}\end{equation}
where $n=l-1$ has been used. Now we can apply (\ref{eq:relation1})
and (\ref{eq:relation2}) on (\ref{eq:relation 3}) to obtain \begin{equation}
\begin{array}{lcl}
a|N,\alpha,\beta\rangle & = & \left\{ \frac{N(\alpha+1)}{(\alpha+\beta+2)}\right\} ^{\frac{1}{2}}\sum_{l=0}^{N-1}\left\{ \frac{(N-1)!(\alpha+2)_{l}(\beta+1)_{N-1-l}}{(\alpha+2+\beta+1)_{N-1}l!(N-1-l)!}\right\} ^{\frac{1}{2}}|l\rangle\\
 & = & \left\{ \frac{N(\alpha+1)}{(\alpha+\beta+2)}\right\} ^{\frac{1}{2}}\sum_{n=0}^{N-1}\sqrt{\omega(n,N-1,\alpha+1,\beta)}|n\rangle,\end{array}\label{eq:gb1}\end{equation}
 where dummy variable $l$ is replaced by $n.$ Therefore, \[
\begin{array}{lcl}
\langle N,\alpha,\beta|a^{\dagger}a|N,\alpha,\beta\rangle & = & \frac{N(\alpha+1)}{(\alpha+\beta+2)}\\
\langle N,\alpha,\beta|a^{\dagger2}a^{2}|N,\alpha,\beta\rangle & = & \frac{N(N-1)(\alpha+1)(\alpha+2)}{(\alpha+\beta+2)(\alpha+\beta+3)}\\
\vdots & \vdots & \vdots\\
\langle N,\alpha,\beta|a^{\dagger l}a^{l}|N,\alpha,\beta\rangle & = & \frac{\left[N(N-1).....(N-l+1)\right]\left[(\alpha+1)(\alpha+2).....(\alpha+l)\right]}{(\alpha+\beta+2)(\alpha+\beta+3).....(\alpha+\beta+l+1)}\\
 & = & \frac{N!(\alpha+l)!(\alpha+\beta+1)!}{(N-l)!\alpha!(\alpha+\beta+l+1)!}\end{array}\]
and \begin{equation}
\begin{array}{lcl}
d_{GBS}(l) & = & \frac{N!(\alpha+l+1)!(\alpha+\beta+1)!}{(N-l-1)!\alpha!(\alpha+\beta+l+2)!}-\left\{ \frac{N(\alpha+1)}{(\alpha+\beta+2)}\right\} ^{l+1}\\
 & = & \frac{\left[N(N-1).....(N-l)\right]\left[(\alpha+1)(\alpha+2).....(\alpha+l+1)\right]}{(\alpha+\beta+2)(\alpha+\beta+3).....(\alpha+\beta+l+2)}-\left\{ \frac{N(\alpha+1)}{(\alpha+\beta+2)}\right\} ^{l+1}\end{array}\label{eq:d(l)broy}\end{equation}
The physical condition $N\geq l+1$ ensures that all the terms in
$d(l)$ are positive. The expression of $d(l)$ is quite complex and
it depends on various parameters (e.g. $\alpha,\,\beta$ and $N$).
Fig 1 shows that for particular values of these parameters HOA is
possible. As it is expected from the earlier works on the properties
of HOA {[}\ref{the:martin}], the depth of nonclassicality is more in
case of $d_{GBS}(9)$ than in $d_{GBS}(8)$. This is consistent with
earlier observation. A systematic study reveals that the probability
of observing HOA increases with the increase of $\alpha$ but it
decreases (i.e the probability of higher order bunching increases)
with the increase of $\beta$. This can be seen clearly in Fig2 and
Fig3. Further it is observed (from Fig1 and Fig2) that for lower
values of $\alpha$ the probability of bunching increases with the
increase of $N$ but for a comparatively large values of $\alpha$
(larger compare to $\beta$) the probability of HOA increases with
the increase in $N$ {[}see Fig 2] but the situation is just opposite
in the case of $\beta$ (as it is seen from Fig3).

\begin{figure}[h]
\centering \scalebox{0.6}{\includegraphics{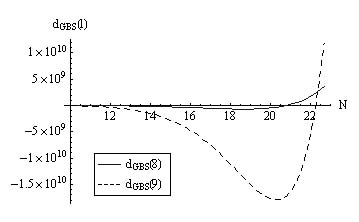}}
\caption{Higher order antibunching can be seen in Generalized
binomial state. Existence of 8th and 9th order antibunching (for
$\alpha=2$ and $\beta=1$)and variation of depth of nonclassicality
with $N$ has been shown.}\label{Fig1}
\end{figure}
\begin{figure}[h]
\centering \scalebox{0.6}{\includegraphics{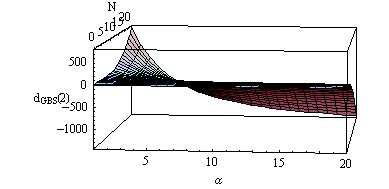}} \label{Fig2}
\caption{Variation of $d_{GBS}(2)$ with $\alpha$ and $N$ for
$\beta$=1.} \label{fig2}
\end{figure}
\begin{figure}[h]
\centering \scalebox{0.6}{\includegraphics{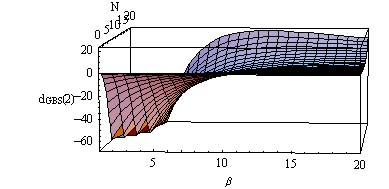}}
\caption{Variation of $d_{GBS}(2)$ with $\beta$ and $N$ for
$\alpha$=10.} \label{Fig3}
\end{figure}
%\begin{figure*}
%\begin{tabular}{ccc}
%\scalebox{0.4}{\includegraphics{gbs3_a2_b1.jpg}}&
%\scalebox{0.3}{\includegraphics{GBS_1.jpg}}&
%\scalebox{0.3}{\includegraphics{GBS_1.jpg}}
%\end{tabular}
%\end{figure*}
While studying different limiting cases of Roy and Roy generalized
binomial state, we have observed that binomial state and number
state always show HOA and $d(l)=0$ for coherent state. This is
consistent with the physical expectation and the conclusions of the
last section.

\section{Other Intermediate States}

As it is mentioned in the earlier sections, there exist several
different intermediate states. For the systematic study of
possibility of observing HOA in intermediate states, we have studied
all the well known intermediate states. Since the procedure followed
for the study of different states are similar, mathematical detail
has not been shown in the subsections below. But from the expression
of $d(l)$ and the corresponding plots it would be easy to see that
the HOA can be observed in all the intermediate states studied
below.

\subsection{Reciprocal binomial state}

Reciprocal binomial state (RBS) can be defined as
{[}\ref{the:reciprocalBS}]
\begin{equation}
|\phi\rangle=\frac{1}{N}\begin{array}{c}
{\scriptstyle N}\\
\sum\\
{\scriptscriptstyle {\scriptstyle k=0}}\end{array}\left(\begin{array}{c}
N\\
K\end{array}\right)^{-1/2}e^{ik(\theta-\pi/2)}|k\rangle\label{eq:rbs1}\end{equation}
where \emph{N} is a normalization constant. Using antinormal
ordering and procedure adapted in section 3, we can
obtain;\begin{equation}
\begin{array}{lcl}
d_{RBS}(l) & = & \begin{array}{c}
{\scriptstyle l+1}\\
\sum\\
i{\scriptstyle =0}\end{array}(-1)^{i}\frac{(l+1)!^{2}}{(l+1-i)!^{2}i!}\frac{(N+(l+1-i\,)!}{N!}-N^{l+1}\\
 & = & \frac{\pi\csc\left(\pi\left(l+N\right)\right)(l+1)!^{2}Gamma(l+1-N)}{M!\left(Gamma(2+l)Gamma(-N)\right)^{2}}-N^{l+1}\end{array}\label{eq:rbs2}\end{equation}
Where $Gamma$ denotes the Gamma function. The possibility of
observing HOA in reciprocal binomial state can be clearly seen from
the Figure 4. But it is interesting to note that the nature of
singularity and zeroes present in the simplified expression of
$d_{RBS}(l)$ as expressed in the last line of (\ref{eq:rbs2}) can
provide us some important information. For example, the underlying
mathematical structure of the criterion of HOA and that of
reciprocal binomial state demands that $l$ and $N$ be integers but
if both of them are integer then $d_{RBS}(l)$ has a singularity as
the $\csc[\pi(l+N)]$ term present in the numerator blows up. But
this local singularity can be circumvented by assuming $l\rightarrow
integer$ and $N\rightarrow integer$. In this situation (i.e. when
$l$ and $N$ tends to integer value) $\langle N^{(l)}\rangle$ is
finite and consequently $d(l)$ is also finite. This is the reason
that the singular nature of the simplified expression of
$d_{RBS}(l)$ is not reflected in the Figure 4. In the analysis of
the $d_{RBS}(l)$ it is also interesting to observe that
$Gamma(-N)=\infty$ for $N=integer$, and in an approximated situation
when $l\rightarrow integer$ and $M=integer$, the $\csc[\pi(l+N)]$
term in the numerator is no more singular and as a result$\langle
N^{(l)}\rangle=0$ and $d_{RBS}=-N^{(l+1)}$. In this situation one
can observe HOA for arbitrarily large values of $l$ and $N$. Thus
physically, it is expected that in reciprocal binomial state higher
order antibunching of any arbitrary order will be seen and HOA will
not be destroyed with the increase of $N$, as it happens (for some
particular values of $\alpha$ and $\beta$) in the case of
generalized binomial state (see Figure 1).
\begin{figure}[h]
\centering
 \scalebox{0.6}{\includegraphics{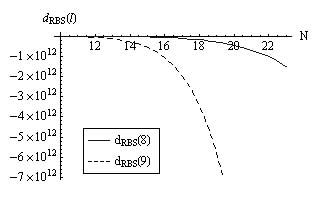}} \caption{Variation of $d_{RBS}(8)$ and $d_{RBS}(9)$ with
photon number $N$ } \label{Fig5}
\end{figure}

\subsection{Negative Binomial State}

Following Barnett {[}\ref{the:Barnett}] we can define Negative
Binomial state (NBS) as

\begin{equation}
|\eta,M\rangle=\sum_{n=M}^{\infty}C_{n}(\eta,M)|n\rangle\label{eq:NBS1}\end{equation} where
$C_{n}(\eta,M)=\left[\left(\begin{array}{c}
n\\
M\end{array}\right)\eta^{M+1}(1-\eta)^{n-M}\right]^{1/2}$, $0\leq\eta\leq1$ and M is a non-negative integer.
This intermediate state interpolates between number state and geometric state. Following the mathematical
techniques adopted in the earlier sections we obtain
\begin{equation}
d_{NBS}(l)=\eta^{-l}\left(\frac{\left(l+M+1\right)!\,_{2}F_{1}\left(-l-1,-l-1;-l-M-1;\eta\right)}{M!}-\frac{(M+1)^{l+1}}{\eta}\right)\label{eq:NBS2}\end{equation}
where $\,_{2}F_{1}(a,b;c;z)$ is a conventional hypergeometric function. Variation of $d(l)$ with various
parameters such as $\eta,\, l$ and $M$ have been studied and are shown in Figure 5- Figure6. From these figures
one can observe that the state is not always antibunched, rather the plot of $d_{NBS}(8)$ has a very sharp rise
near $\eta\approx.15$ and $M\approx10$. From Figure 5 we can observe that the possibility of higher order
bunching get destroyed with the increase of $M$ or $\eta$. The second feature has clearly been shown in Figure
6, where we can see that $d_{NBS}(8)$ become positive for a small region for $M=10$ but for a bigger region
($\eta\approx.3$ to $\eta=1$) it remains negative and thus shows the existence of HOA. The broad features remain
same for the other orders (other values of $l$) of antibunching.
\begin{figure}[h]
\centering
 \scalebox{0.5}{\includegraphics{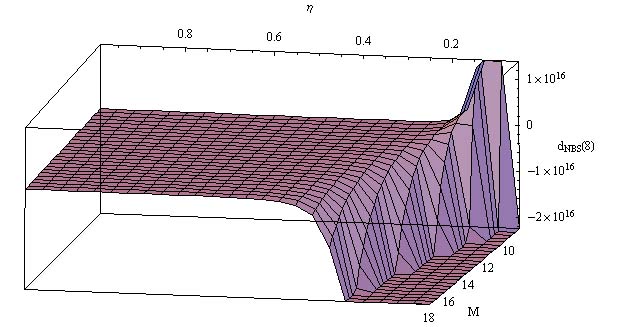}} \caption{Variation of $d_{NBS}(8)$  with $\eta$ and $M$
} \label{Fig6}
\end{figure}

\begin{figure*}[h]
\centering
\begin{tabular}{cc}
 \scalebox{0.5}{\includegraphics{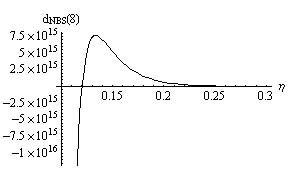}}& \scalebox{0.5}{\includegraphics{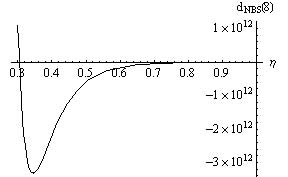}}
\end{tabular}\caption{Variation of
$d_{NBS}(8)$  with $\eta$ for $M=10$ }\label{Fig7}
\end{figure*}

In the limit $N\rightarrow0$ the negative binomial state reduces to
geometric state (GS). In this limit $d_{NBS}(l)$ reduces to
\begin{equation}
d_{GS}(l)=\frac{1}{\eta^{l+1}}\left((1-\eta)^{l+1}\eta(l+1)!-1\right).\label{eq:geometric
1}\end{equation} It is interesting to observe that the above
expression has a singularity at $\eta=0$ and
$d_{GS}(l)\rightarrow-\infty$ in the limit $\eta\rightarrow 0$.
Consequently one negative values of $d_{GS}$ can be seen at very
very small values $\eta$ but this is not the signatures of HOA,
rather this is the signature of the existence of a strong
singularity in the neighborhood. This can further justified by the
fact that for any finite value of $l$ there does not exist any real
root, (whose value is close to zero or which is negligibly small
compared to 1) of $d_{GS}(l)=0$. Thus there is no oscillation
between bunching and antibunching. We further observe that
$d_{GS}(l)\rightarrow-1$ in the limit $\eta\rightarrow1$ and for
$l\geq3$ there exists only one physically acceptable real root of
$d_{GS}(l)=0$. By physically acceptable real root we mean that it
lies in $[0,1]$. Before this value of $\eta$ (or before the
physically acceptable real root) the state shows higher order super
poissonian photon statistics but immediately after the root it
becomes negative and thus shows HOA or higher order subpoissonian
photon statistics. As we increase $l$ the real root shifts in right
side of the real axis (i.e towards $\eta=1$). It can be clearly seen
in the Figure 7. From this figure it can be easily seen that it
satisfies all the physical properties of HOA derived in
{[}\ref{the:martin}].
\begin{figure*}[h]
\centering
 \scalebox{0.7}{\includegraphics{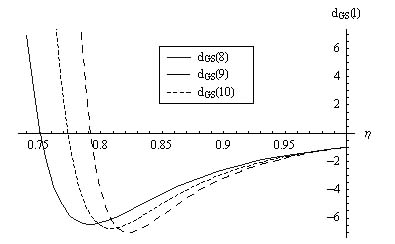}} \caption{Variation of $d_{GS}(8)$, $d_{GS}(9)$ and
$d_{GS}(10)$ with respect to $\eta$. \label{Fig4}}
\end{figure*}

\subsection{Photon added coherent state}

Photon added coherent state (\ref{eq:photonadded1}) or PACS, which
was introduced by Agarwal and Tara {[}\ref{the:agarwal-photonadded}]
can be defined as
\begin{equation}
|\alpha,m\rangle=\frac{\exp(-\frac{|\alpha|^{2}}{2})}{\sqrt{L_{m}(-|\alpha|^{2})m!}}\sum_{n=0}^{\infty}\alpha^{n}\,\frac{\sqrt{(m+n)!}}{n!}|n+m\rangle\label{eq:photonadded1n}\end{equation}
where $L_{m}(x)$ is Lauguere polynomial of $mth$ order. Rigorous operator algebra yields

\begin{equation}
\begin{array}{lcl}
d_{PACS}(l) & = & \frac{\exp(-\alpha^{2})\alpha^{2l+2}((l+m+1)!)^{2}\,_{P}F_{Q}\left(\left\{ 1,2+l+m,2+l+m\right\} ;\left\{ 2+l,2+l,m+1\right\} ;\alpha^{2}\right)}{(m!(l+1)!)^{2}\,_{1}F_{1}\left(-m;1;-\alpha^{2}\right)}\\
 & - & \left(\frac{\exp(-\alpha^{2})\left(-m+m\,_{1}F_{1}\left(1+m;1;\alpha^{2}\right)+(1+m)\alpha^{2}\,_{1}F_{1}\left(2+m;2;\alpha^{2}\right)\right)}{\,_{1}F_{1}\left(-m;1;-\alpha^{2}\right)}\right)^{l+1},\end{array}\label{eq:photonadded2}\end{equation}
where, $\,_{P}F_{Q}$ is the generalized Hypergeometric function. The
analytic expression for $d_{PACS}(l)$ is quite complicated and it is
difficult to conclude anything regarding its photon statistics
directly from (\ref{eq:photonadded2}) but we have investigated the
variation of $d_{PACS}$ with $\alpha,\, l$ and $m$ and could not
find any region which does not show HOA. Therefore, HOA can be seen
in this particular intermediate state. This fact is manifested in
Fig 8 and Fig 9. From these figures it is easy to observe that depth
of nonclassicality increases monotonically with the increase of $m$
and $l$. The variation of depth of nonclassicality with $\alpha$ has
a deep for a small value of $\alpha$ (see Fig 8. and Fig. 9).
Although $d_{PACS}$ is always negative, initially its magnitude
($d_{PACS}(l)$ without the negative sign) increases, then decreases
and then become a monotonically increasing function. Actually for
the smaller values of $\alpha$, an effective contribution from the
combination of all the hypergeometric functions appears and
dominates but as soon as $\alpha$ increases a bit, the
$\exp(-\alpha^{2})$ term starts dominating and as a consequence
depth of nonclassicality increases monotonically. Here we would also
like to note that in contrast to the photon added coherent state,
$d(l)$ is always positive for the analogous state
$|\alpha,-m\rangle$, introduced by Sivakumar
{[}\ref{the:Shiavakumar}]. Thus $|\alpha,-m\rangle$ always shows
higher order superpoisonian photon statistics. Further, we would
like to note that photon added coherent state which, is intermediate
between coherent and fock state has already been experimentally
generated in 2004 {[}\ref{the:photonadded-experiment}]. Therefore,
it is technically feasible to observe higher order antibunching for
an intermediate state.

\begin{figure}[h]
\centering
 \scalebox{0.6}{\includegraphics{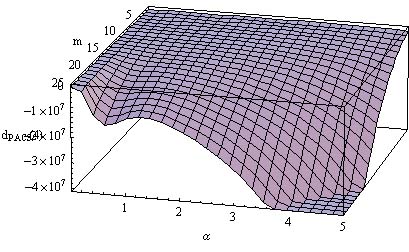}} \caption{Variation of $d_{PACS}(4)$  with
 $\alpha$ and $m$} \label{Fig8}
\end{figure}
\begin{figure}[h]
\centering
 \scalebox{0.6}{\includegraphics{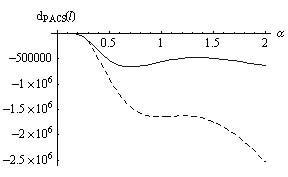}} \caption{Variation of $ 10 d_{PACS}(3)$ and $d_{PACS}(4)$ with
 $\alpha$ for $m=15$, the solid line denotes $d_{PACS}(3)$ and the dashed line denotes $d_{PACS}(4)$ to keep
 show the variation in the same scale $d_{PACS}(3)$ is multiplied by 10. The plot shows that depth of
 nonclassicality of $d_{PACS}(4)$ is always greater than that of $d_{PACS}(3)$ which is consistent with the
 properties of HOA} \label{Fig9}
\end{figure}
\subsection{Hypergeometric state}

Following {[}\ref{the:hypergeomtric}] hypergeometric state (HS) can
be defined as \begin{equation}
|L,M,\eta\rangle=\sum_{n=0}^{M}H_{n}^{M}(\eta,L)|n\rangle\label{eq:hyper1}\end{equation}
where the probability $\eta$ is a real parameter satisfying
$0<\eta<1.$ $L$ is a real number satisfying \begin{equation} L\geq
max\left\{ M\eta^{-1},\, M(1-\eta)^{-1}\right\}
,\label{eq:hyper2}\end{equation}
\begin{equation}
H_{n}^{M}(\eta,L)=\left[\left(\begin{array}{c}
L\eta\\
n\end{array}\right)\left(\begin{array}{c}
L\left(1-\eta\right)\\
M-n\end{array}\right)\right]^{\frac{1}{2}}\left(\begin{array}{c}
L\\
M\end{array}\right)^{-\frac{1}{2}},\label{eq:hyper3}\end{equation} and \begin{equation} \left(\begin{array}{c}
\alpha\\
n\end{array}\right)=\frac{\alpha(\alpha-1)...(\alpha-n+1)}{n!},\,\,\,\,\,\,\,\left(\begin{array}{c}
\alpha\\
0\end{array}\right)\equiv1.\label{eq:hyper4}\end{equation} Here
$\alpha$ is not necessarily an integer. Using the techniques adopted
in the earlier sections and a bit of operator algebra we can obtain
a closed form analytic expression for $d(l)$ as \begin{equation}
d_{HS}(l)=-\left(M\eta\right)^{l+1}+\frac{(L-l-1)!M!(L\eta)!}{L!(M-l-1)!(L\eta-1-l)!}\label{eq:hyper5}\end{equation}
From Fig. 10 it is clear that HOA can be observed in hypergeometric
state. It is also observed that the depth of nonclassicality
increases with the increase in $\eta$ and $M$. Hypergeometric state
reduces to binomial state, coherent state, number state and vacuum
state in different limits of $M,$ $L$ and $\eta$. It has been
verified that if we impose those limits on $d_{HS}$ then we obtain
corresponding photon statistics.

\begin{figure}[h]
\centering
 \scalebox{0.6}{\includegraphics{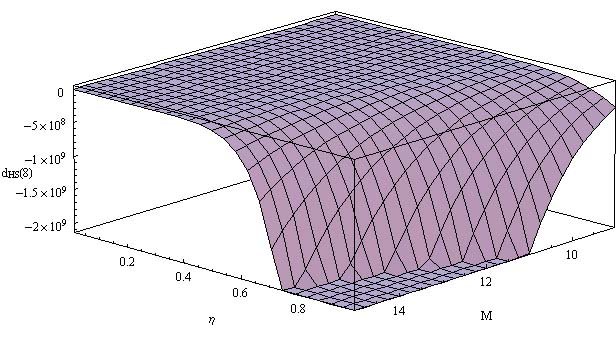}} \caption{Variation of $d_{HS}(8)$  with
 $\eta$ and $M$, when lowest allowed values of $L$ have been chosen at every point. } \label{Fig10}
\end{figure}

\section{Conclusions}

In essence all the intermediate states studied in the present paper
\textcolor{black}{show HOA i.e. higher order subpoissonian photon
statistics. But it neither mean that all the intermediate states are
higher order antibunched (for example $|\alpha,-m\rangle$ is always
higher order bunched) nor an intermediate state which shows HOA will
show it for all possible values of the control parameters (for
example negative binomial state and generalized binomial state shows
both higher order bunching and higher order antibunching for various
parametric values). Thus we can conclude that, as far as HOA is
concerned there does not exist any common characteristics among the
different intermediate states but most of them show HOA. Further, we
have seen from (Figure1-Figure10) that the depth of nonclassicality
of a higher order antibunched state varies with different control
parameters (e.g. $\alpha$, $N$, $m$ etc.). These parameters
represent some physical quantity and their value may be controlled
and consequently by controlling these parameters we can control the
depth of nonclassicality.
 Photon statistics (factorial moment) of an intermediate state
can be obtained experimentally by using homodyne detection (photon
counting) technique. These facts along with the recent success in
experimental production of intermediate state open up the
possibility of experimental observation of HOA in intermediate
state. Thus the present work strongly establishes the fact that HOA
is not a rare phenomenon. }

\textcolor{black}{The prescription followed in the present work is easy and straight forward and it can be used
to study the possibilities of observing higher order antibunching in other intermediate states (such as negative
hyper geometric state, excited binomial state and odd excited binomial state) and other physical systems. Thus
it opens up the possibility of studying higher order nonclassical effects from a new perspective. This is also
important from the application point of view because any probabilistic single photon source used for quantum
cryptography has to satisfy the condition for higher order antibunching. }

\textbf{Acknowledgement}: AP thanks to DST, India for partial financial
support through the project grant SR\textbackslash{}FTP\textbackslash{}PS-13\textbackslash{}2004.

\end{document}